\documentclass[]{article}
\usepackage{graphics}
\usepackage{dcolumn}
\usepackage{bm}

\begin{document}

\title{\bf Research of Gravitation in Flat Minkowski Space}

\author{Kostadin Tren\v{c}evski$^1$, Emilija G. Celakoska$^2$ and Vladimir Balan$^3$\bigskip \\
$^1$Faculty of Natural Sci. and Math., Ss. Cyril and Methodius Univ., \\
P.O.Box 162, Skopje, Macedonia, e-mail: kostatre@pmf.ukim.mk\\
$^2$Faculty of Mechanical Engineering, Ss. Cyril and Methodius Univ.,\\
P.O.Box 464, Skopje, Macedonia, e-mail: cemil@mf.edu.mk \\
$^3$Politehnica Univ. of Bucharest, Splaiul Independentei 313,\\
RO-060042 Bucharest, Romania, e-mail: vbalan@mathem.pub.ro}

\date{}

\maketitle

\begin{abstract}
In this paper it is introduced and studied an alternative theory of gravitation
in flat Minkowski space.
Using an antisymmetric tensor $\phi$, which
is analogous to the tensor of electromagnetic field, a non-linear
connection is introduced. It is very convenient for studying the
perihelion/periastron shift, deflection of the light rays near the
Sun and the frame dragging together with geodetic precession, i.e.
effects where angles are involved. Although the corresponding
results are obtained in rather different way, they are the same as
in the General Relativity. The results about the barycenter of two
bodies are also the same as in the General Relativity. Comparing the
derived equations of motion for the $n$-body problem with the
Einstein-Infeld-Hoffmann equations, it is found that they differ
from the EIH equations by Lorentz invariant terms of order $c^{-2}$.
\medskip


\end{abstract}
\maketitle

\vspace{-0.2cm}
\section{Introduction}
\label{intro}

In this paper, the gravitational phenomena are studied in flat Minkowski space and
this approach is a small step ahead of the
Special Relativity. In the literature there are some attempts the results of the General
Relativity to be obtained in flat space-time and a study of such
attempts and a proposed theory is given in \cite{B8}. Another
example is the teleparallel approach \cite{AlPe}, where the metric
is hidden in the frame. Teleparallel gravity is reduced to General
relativity and therefore calculations for the gravitational tests
are not necessary. However, the study in this paper is broader and we also make the
calculations (up to $c^{-2}$) to investigate the agreement with the
basic gravitational tests.

For the equations of motion the position of the observer is also
important, i.e. whether he is away from the gravitational field, or
inside the gravitational field.
Indeed, the equations depend only on the chosen coordinate
system, but the parameters in the equations depend on the position
of the observer in its local coordinate frame. Such position
dependent parameters are for example the acceleration toward the
gravitational bodies. So, we can distinguish four cases:

1. The observer is far from gravitation and the coordinates are
orthonormal;

2. The observer is inside the gravitational field and the
coordinates are ordinary (curvilinear);

3. The observer is inside the gravitational field and the
coordinates are orthonormal;

4. The observer is far from gravitation and the coordinates are
ordinary (curvilinear).

In this paper, we focus on the case 1. Specially, the theory will be
covariant with respect to the Lorentz transformations with constant
elements, analogously to the Special Relativity, because of the
freedom of the choice of the inertial coordinate system far from
gravitation, where the observer is placed. Cases 2 and 3 are 
more complicated and the case 3 is considered in 
\cite{TC3}. The case 4 is a subject of the General Relativity (GR),
more precisely the Einstein-Infeld-Hoffmann equations,
which will be supported in 7.6.

In section 2 we present a nonlinear connection in the Minkowskian space.
Such a research
offers a great convenience in calculations which have been used as advantage also
in some other approaches \cite{AP,N}, etc. It gives a very close
relationship with the electrodynamics (section 2) which gives
possibility for quantization of gravitation and unification with the
other interactions, since the other interactions are considered in
flat space (see \cite{B3} for interesting discussions on the topic).

Non-linear connections are widely used at present time. For example,
non-linear connections using Finsler geometry are studied in
\cite{B4,B5,B6} and also in \cite{B7,B9,B10,B11,B12}. But although
for studying gravitation both nonlinear connection and research in
flat space are not new (\cite{Log}),
in this paper we propose an approach obeying both characteristics.

We use $ict$ convention (see pp. 51 in \cite{M} about $ct/ict$
conventions). So, we work with the Euclidean metric
$\hbox{diag}(1,1,1,1)$ and upper and lower indices will not differ.

\vspace{-0.2cm}
\section{Introduction of a Non-linear Connection}
\label{sec:2}

Firstly, we explain why it is not convenient to use linear
connection.
Let us examine the effect of
a linear connection concerning the 4-velocities. The parallel
transport of any vector in the direction of a 4-vector of velocity
$(V_1,V_2,V_3,V_4)$ means that parallel transport is made in each of
the four directions $(1,0,0,0)$, $(0,1,0,0)$, $(0,0,1,0)$ and
$(0,0,0,1)$, these are multiplied by $V_1$, $V_2$, $V_3$, and $V_4$
respectively and then all is added together. But, we can not
consider a 4-velocity as a translation, since that is not supported
by the special-relativistic addition. Rather, a 4-velocity should be
regarded as a Lorentz transformation with its incorporated
hyperbolic properties in the 4-dimensional space. So, we will
consider a non-linear connection in a sense that the condition
$\nabla_{aX+bY}=a\nabla_X+b\nabla_Y$ is dismissed. The construction
will be made in three steps.

\vspace{-0.2cm}
\subsection{Using an analogy from electromagnetism}
\label{sec:2-1}

Firstly, we will make a complete analogy with the electromagnetism,
where instead of the charge $e$ we will consider mass $M$, and
instead of the potential $\frac{e}{r}$ we will consider the
gravitational potential $\frac{GM}{r}$, assuming that $M$ has the
same value in each inertial coordinate system. Further we will
introduce an antisymmetric tensor analogous to the tensor of
electromagnetic field. We accept a priori that the velocity of the
gravitational interaction is $c$, which would enable us to find this
tensor when the source of gravitation field is accelerated.

Let us consider the motion of a test body under the influence of a
gravitational body with mass $M$ concentrated into a point with a
time dependent 4-vector of velocity
$$(U_{1},U_{2},U_{3},U_{4}) = \frac{1}{\sqrt {1-u^{2}/c^{2}}} \Bigl (
\frac{u_{x}}{ic}, \frac{u_{y}}{ic}, \frac{u_{z}}{ic}, 1\Bigr )
,\eqno{(2.1)}$$ where $\vec{u}=(u_{x},u_{y},u_{z})$ is the
corresponding 3-vector of velocity. Assume that the 4-vector of
velocity of a test body with mass $m$ is given by
$$ (V_{1},V_{2},V_{3},V_{4}) = \frac{1}{\sqrt {1-v^{2}/c^{2}}}\Bigl (
\frac{v_{x}}{ic}, \frac{v_{y}}{ic}, \frac{v_{z}}{ic}, 1\Bigr )
.\eqno{(2.2)}$$

We build an antisymmetric tensor field $\phi_{ij}$, in the following
way. Firstly, we consider a special case when the sources of
gravitation move with constant velocities. It is sufficient to
define this tensor for a stationary body with point mass $M$. Then,
using the Lorentz transformations and the principle of superposition
of the fields, the tensor is theoretically well defined in this
special case. In this case, at the point $(x,y,z)$ $\phi$ is
defined by
$$(\phi_{ij})=\left[ \begin{array}{cccc}
0 & 0 & 0 & \frac{GM}{r^3c^2}(x-x_0)\\[1.9ex]
0 & 0 & 0 & \frac{GM}{r^3c^2}(y-y_0)\\[1.9ex]
0 & 0 & 0 & \frac{GM}{r^3c^2}(z-z_0)\\[1.9ex]
-\frac{GM}{r^3c^2}(x-x_0) & -\frac{GM}{r^3c^2}(y-y_0) &
-\frac{GM}{r^3c^2}(z-z_0) & 0\end{array}\right ] , \eqno{(2.3)}$$
where $(x_0,y_0,z_0)$ is the position of the gravitational body.

The 3-vector $c^2(\phi_{41},\phi_{42},\phi_{43})$ is the Newton
acceleration toward the gravitational body, which is analogous to
the electric field $\vec{E}$. The physical interpretation of the
components $\phi_{ij}$ for $1\le i,j\le 3$ will be given by (2.17).

Notice that using this tensor in flat Minkowski space it is obtained
a general formula for frequency redshift/blueshift \cite{TC}, which
simultaneously explains the Doppler effect, gravitational redshift and under one
cosmological assumption it also explains the cosmological redshift and the
blueshift arising from the Pioneer anomaly. The gravitational redshift there
is a consequence of the attraction force near the gravitational bodies and
we do not need curved space any more.

Now let us consider arbitrary time variable vector $\vec{u}$ of the
source of gravitation. Analogously as obtaining Lienard-Wiechert
potentials in electrodynamics, the components of the tensor in case
of gravitation can be obtained at each space-time point, using that
the gravitational interaction transmits with velocity $c$. So, we
get the following analogous formulae as in electrodynamics
$$c^2(\phi_{41},\phi_{42},\phi_{43})=
-\frac{GM}{(R-\frac{\vec{R}\cdot \vec{u}}{c})^3} \Bigl
(\vec{R}-\frac{\vec{u}}{c}R\Bigr )-$$ $$-\frac{GM}{c^2
(R-\frac{\vec{R}\cdot \vec{u}}{c})^3}\vec{R}\times \Bigl [ \Bigl
(\vec{R}-\frac{\vec{u}}{c}R\Bigr )\times \dot{\vec{u}}\Bigr ],
\eqno{(2.4a)}$$
$$\frac{c}{i}(\phi_{32},\phi_{13},\phi_{21})=\frac{1}{R}\vec{R}\times
(\phi_{41},\phi_{42},\phi_{43}).\eqno{(2.4b)}$$ Here $\vec{u}$ is
the velocity of the gravitational body, $\vec{R}$ is the 3-vector
from the gravitational body to the considered point $(x,y,z,ict)$ in
the chosen coordinate system calculated at the space-time point
$(x',y',z',ict')$ of the gravitational body, such that after time
$t-t'$ of transmission of the interaction, it arrives at the
considered point $(x,y,z,ict)$. Thus, $t'$ appears as a solution of
the equation
$$t=t'+\frac{R(t')}{c}.\eqno{(2.5)}$$
In (2.4a) $\dot{\vec{u}}=\partial \vec{u}/\partial t'$ and $R=\vert
\vec{R}\vert$.

In the special case when $\dot{\vec{u}}=0$ the equation (2.4a)
reduces to
$$c^2(\phi_{41},\phi_{42},\phi_{43})=
-\frac{GM}{R^3}\vec{R}\frac{1-\frac{u^2}{c^2}} {\Bigl
(1-\frac{u^2}{c^2}\sin^2\theta \Bigr )^{3/2}},\eqno{(2.6)}$$ where
$\theta$ is the angle between $\vec{R}$ and $\vec{u}$, and $\vec{R}$
is the 3-vector from the gravitational body to the considered point
at time $t$. This special case can be deduced directly from (2.3)
using the Lorentz transformations.

In this paper we will work up to $c^{-2}$ approximation. Since for
the 2-body problem $\vec{R}$ is collinear with $\dot{\vec{u}}$, the
last term in (2.4a) can be neglected for $c^{-2}$ approximation.
Hence, in this paper we can use the equality (2.6), except in
section \ref{sec:8}, where the $n$-body problem is considered.

A natural question appears about the analog of the 4-vector
potential from the electromagnetism. It is treated in some previous
papers \cite{T1,T5} and it is not necessary to consider it in this
paper.

We can resume, so far, that in case of gravitation we accepted some
facts from the electromagnetism. But we must emphasize that there
are two essential differences, which will be considered in the
subsections 2.2 and 2.3. The gravity is associated with a spin-2
field rather than the spin-1 field of the electromagnetism.

i) While the charge $e$ in electrodynamics is invariant scalar in
all coordinate systems, the gravitational mass $M$ is not invariant.
Since the inertial mass is not Lorentz invariant according to the
Special Relativity, it is naturally to expect that the gravitational
mass is not invariant in flat Minkowski space.
Thus, the tensor
$\phi$ must be modified, and this will be done in 2.2.

ii) The equations of motion can not simply copy the Lorentz force
from the electrodynamics, because it gives a parallel transportation
only of a single vector, the 4-vector of velocity, but not of an
arbitrary vector. Thus, in case of gravitation we must modify the
Lorentz force, and it will be done in 2.3. 
Moreover, while the
Lorentz force acting on the charged particles depends on the
electromagnetic field at the considered point and not on the
velocity of the source of the electromagnetic field, in case of
gravitation, as we shall see, the motion depends on the source of
gravitation very explicitly. This dependence in GR is implicitly
contained in the Einstein's equations and it is explicitly visible
in the Einstein-Infeld-Hoffmann equations.

\vspace{-0.2cm}
\subsection{Influence of the masses to the gravitational force
and acceleration}

A mass far from gravitation measured by an observer far from
gravitation will be called proper mass and will be denoted by $m$,
$M$, $m_1$, $m_2$,... An observer far from gravitation observing a
body with proper mass $m$ that has fallen into a gravitational field
with gravitational potential $\frac{GM}{R}$, will measure
$\frac{m}{1+{\frac{G M}{Rc^2}}}$ for the mass of the body. It is
convenient the scalar $\mu =1+\frac{GM}{Rc^2}$ to call also
gravitational potential.
Assume that the test body has a small mass $m$ with respect to the
gravitational body. Then this is in accordance with the preserving
of the energy in a gravitational field, such that considering also
the kinetic energy, the mass $\frac{m}{1+\frac{G M}{Rc^2}}
\frac{1}{\sqrt{1-\frac{v^2}{c^2}}}$ will be unchanged up to $c^{-2}$
during the motion of the test body.

Let us consider two bodies with masses $m_1$ and $m_2$ on a distance
$R$ between their centers. Then the mass $m_1$
is observed to be $\frac{m_1}{1+\frac{Gm_2}{Rc^2}}$
under the influence of the other mass $m_2$, and the mass
$m_2$ is observed to be $\frac{m_2}{1+\frac{Gm_1}{Rc^2}}$ under the influence of the other
mass $m_1$. So, the gravitational force which acts
on the body with mass $m_1$ and is caused by the body with mass
$m_2$ is given by
$$\vec{f}=\frac{m_1}{1+\frac{Gm_2}{Rc^2}}\nabla
\frac{Gm_2}{R(1+\frac{G m_1}{Rc^2})},\eqno{(2.7)}$$ while the
acceleration of the body with mass $m_1$ is assumed to be
$$\vec{a}=\frac{1}{1+\frac{Gm_2}{Rc^2}}\nabla
\frac{Gm_2}{R(1+\frac{G m_1}{Rc^2})}.\eqno{(2.8)}$$ The formulae
(2.7) and (2.8) will be
generalized below by (2.10) and (2.11). We must emphasize that these
formulae are given with respect to an observer far from gravitation,
assuming also that the observer does not move with respect to the
gravitational bodies. Here, the distance $R$ is a function of 6
coordinates: 3 coordinates of the body with mass $m_1$ and 3
coordinates of the body with mass $m_2$, and the gradient is taken
with respect to the coordinates of the body with mass $m_1$. It is
easy to see that up to $c^{-2}$ the acceleration (2.8) can be
written in the form
$$\vec{a}=-\frac{\vec{R}}{R}\frac{Gm_2}{R^2}
\Bigl (1-\frac{G(2m_1+m_2)}{Rc^2}\Bigr ),\eqno{(2.9)}$$ where
$\vec{R}$ is the vector from the body with mass $m_2$ towards the
body with mass $m_1$.

In section \ref{sec:8} we will consider the general equations for
$n$-body problem. Then it will be necessary to use a more general
formula for the acceleration. If we consider the interaction of two
bodies, for example with masses $m_1$ and $m_2$, we must use that
their masses in the gravitational field are
$$m_1/\Bigl [(1+\frac{Gm_2}{r_{12}c^2})(1+\frac{Gm_3}{r_{13}c^2})
...\Bigr ]\quad \hbox{and} \quad m_2/\Bigl
[(1+\frac{Gm_1}{r_{21}c^2})(1+\frac{Gm_3}{r_{23}c^2}) ...\Bigr ]$$
respectively, where $r_{ij}$ is the distance between the bodies with
masses $m_i$ and $m_j$. Now, analogously to (2.7), and (2.8) for the
force/acceleration of the body with mass $m_1$ caused by the mass
$m_2$ we accept {\em axiomatically} that
$$\vec{f}=\frac{m_1}{(1+\frac{Gm_2}{r_{12}c^2})(1+\frac{Gm_3}{r_{13}c^2})
...} \nabla
\frac{Gm_2}{r_{12}(1+\frac{Gm_1}{r_{12}c^2})(1+\frac{Gm_3}{r_{23}c^2})
...},\eqno{(2.10)}$$
$$\vec{a}=\frac{1}{(1+\frac{Gm_2}{r_{12}c^2})(1+\frac{Gm_3}{r_{13}c^2})
...} \nabla
\frac{Gm_2}{r_{12}(1+\frac{Gm_1}{r_{12}c^2})(1+\frac{Gm_3}{r_{23}c^2})
...}.\eqno{(2.11)}$$ Analogously to (2.9), in this general case we
obtain
$$\vec{a}=\Bigl [\Bigl (1+\frac{Gm_2}{r_{12}c^2}
\Bigr )\Bigl (1+\frac{Gm_1}{r_{12}c^2}\Bigr )^2 \Bigl
(1+\frac{Gm_3}{r_{13}c^2}\Bigr )\times$$ $$ \times\Bigl
(1+\frac{Gm_3}{r_{23}c^2}\Bigr ) \Bigl
(1+\frac{Gm_4}{r_{14}c^2}\Bigr ) \Bigl
(1+\frac{Gm_4}{r_{24}c^2}\Bigr )...\Bigr ]^{-1} \nabla
\frac{Gm_2}{r_{12}}.\eqno{(2.12)}$$ Notice that according to the
assumptions that the observer is far from gravitation and the
gravitational bodies do not move with respect to the observer, the
acceleration from 2.1 is given by $\nabla
\displaystyle\frac{Gm_2}{r_{12}}$. Thus, for moving bodies with
respect to the observer, up to $c^{-2}$, the components
$\phi_{14},\phi_{24},\phi_{34},\phi_{41},\phi_{42},\phi_{43}$ should
be multiplied by the coefficient in front of $\nabla\displaystyle
\frac{Gm_2}{r_{12}}$ in (2.12). Since the components $w_x,w_y,w_z$
are much smaller than $a_x,a_y,a_z$, we can conclude that {\em all
the components of the tensor $\phi$ should be multiplied by the
coefficient in front of $\nabla\displaystyle \frac{Gm_2}{r_{12}}$
in} (2.12). This coefficient in (2.12) is a scalar in the
Minkowskian space up to $c^{-2}$, and hence the product, i.e. the
modified tensor $\phi$, will preserve its tensor character. We agree
that further on, $\phi$ will always mean this modified tensor.

We shall draw some conclusions. For example, if $m_1$ is negligible
small mass and $m_2=M$ is non-zero mass of a stationary body, then
the acceleration of the body with mass $m_1$ is equal to
$\vec{a}=-\frac{\vec{R}}{R}\frac{\frac{G M}{R^2}}{1+\frac{G
M}{Rc^2}} $. This acceleration can be written as
$$\vec{a}=c^2\nabla \ln \Bigl (1+\frac{GM}{Rc^2}\Bigr ).
\eqno{(2.13)}$$ Now, it is clear that the potentials $\mu
=1+\frac{GM}{Rc^2}$ and $C\mu $, where $C$ is a constant, lead to
the same acceleration.

At the end of this subsection we emphasize the following remark
about the preserving the energy of a system of $n$-bodies with
masses $m_1,m_2,...,m_n$. According to the accepted change of the
mass near gravitational bodies, the energy of the $i$-th body,
including the energy in rest $m_ic^2$ is equal to
$$\frac{m_ic^2}{\sqrt{1-\frac{v_i^2}{c^2}}
\prod\limits_{j\neq i}(1+\frac{Gm_j}{r_{ij}c^2})}.$$ Following the
electrodynamics analogy, as the density of energy caused by charged
particles is given by $\displaystyle\frac{E^2+H^2}{8\pi}$, in case
of gravitation we have that the density of energy is given by
$\displaystyle\frac{\vec{a}^2+\vec{w}^2c^2}{8\pi G}$, because
$c\vec{w}$ corresponds to the magnetic field (see (2.17)). Since
$w\sim c^{-2}$, this energy density can be approximated by
$\displaystyle\frac{\vec{a}^2}{8\pi G}$. Hence the total energy is
given by
$$\sum_{i=1}^n\frac{m_ic^2}{\sqrt{1-\frac{v_i^2}{c^2}}
\prod\limits_{j\neq i}(1+\frac{Gm_j}{r_{ij}c^2})}+ \frac{1}{8\pi
G}\int \vec{a}^2dV. \eqno{(2.14)}$$ Using that $\frac{1}{8\pi G}\int
\vec{a}^2dV = \sum\limits_{i,j,j\neq
i}\frac{Gm_im_j}{2r_{ij}}+const.$, we obtain that up to a constant
summand the total energy can be written in the form
$$\sum_{i=1}^n\frac{m_ic^2}{\sqrt{1-\frac{v_i^2}{c^2}}}-
\frac{1}{8\pi G}\int \vec{a}^2dV.$$ This formula is the same as in
GR, and hence the conclusion in GR that the density of energy is
$-\frac{\vec{a}^2}{8\pi G}$ \cite{L}, instead of
$\frac{\vec{a}^2}{8\pi G}$. Although this energy $\int
\frac{\vec{a}^2}{8\pi G}dV$ and also the kinetic energy take part in
determining the barycenter of the system of bodies, both energies do
not contribute to the acceleration of the other bodies. Only the
mass $m_i/\prod\limits_{j\neq i}(1+\frac{Gm_j}{r_{ij}c^2})$ plays
role in the acceleration towards the $i$-th body. This is visible
from (2.10) and (2.11). Also notice that the barycenter of the
bodies remains unchanged, compared with the GR. Namely, analogously
as obtaining the barycenter in the GR and in electrodynamics
\cite{L}, in this case one obtains again the same radius-vector
$$\vec{r}_b=
\frac{\sum\limits_{i=1}^n\vec{r}_i \Bigl
(m_ic^2+\frac{1}{2}m_iv_i^2-\frac{Gm_i}{2}\sum\limits_{j\neq
i}\frac{m_j}{r_{ij}}\Bigr )} {\sum\limits_{i=1}^n \Bigl
(m_ic^2+\frac{1}{2}m_iv_i^2-\frac{Gm_i}{2}\sum\limits_{j\neq
i}\frac{m_j}{r_{ij}}\Bigr )}. \eqno{(2.15)}$$

\vspace{-0.2cm}
\subsection{Equations of parallel displacement}

Notice that in a system of four orthonormal vectors $A_{i1}$,
$A_{i2}$, $A_{i3}$ and $A_{i4}$, where $A_{i\alpha}$ is the $i$-th
coordinate of the $\alpha$-th vector, using that $A_{i\alpha}$ is an
orthogonal matrix, i.e. $AA^T=I$, the following tensor
$$\frac{dA_{i\alpha}}{ds}A_{j\alpha},\eqno{(2.16)}$$
$ds=ic\sqrt{1-\frac{v^2}{c^2}}dt$ is also skew-symmetric as
$\phi_{ij}$ is. The formula (2.16) is invariant under the linear
transformation $A_{i\alpha}\rightarrow
B_{i\alpha}=A_{i\beta}R_{\beta \alpha}$, where $R$ is an orthogonal
matrix with constant elements. In the special case when $U_i=V_i$,
we assume that the two tensors, $\phi_{ij}$ and the tensor in
(2.16), are equal. Then the physical interpretation of $\phi_{ij}$
can be obtained using the tensor (2.16). Since (2.16) is invariant
under the linear transformation $A\rightarrow AR$, without loss of
generality, we may assume that $A_{ij}=\delta_{ij}$ at the
considered point, and hence the components of (2.16) are 3-vector of
acceleration and 3-vector of angular velocity. We represent $\phi$
in the following form
$$ \phi = \left [\matrix{
0 & -i\omega _{z}/c & i\omega _{y}/c & -a_{x}/c^{2}\cr i\omega
_{z}/c & 0 & -i\omega _{x}/c & -a_{y}/c^{2}\cr -i\omega _{y}/c &
i\omega _{x}/c & 0 & -a_{z}/c^{2}\cr a_{x}/c^{2} & a_{y}/c^{2} &
a_{z}/c^{2} & 0\cr }\right ],\eqno{(2.17)}$$ where
$\vec{a}=(a_{x},a_{y},a_{z})$ is the 3-vector of acceleration and
$\vec{w}= (w_{x},w_{y},w_{z})$ is the 3-vector of angular velocity.
Indeed we accept the following notations
$$a_x=\phi_{41}c^2=-\phi_{14}c^2,\; a_y=\phi_{42}c^2=-\phi_{24}c^2,\;
a_z=\phi_{43}c^2=-\phi_{34}c^2,$$
$$w_x=ic\phi_{23}=-ic\phi_{32},\;
w_y=ic\phi_{31}=-ic\phi_{13},\; w_z=ic\phi_{12}=-ic\phi_{21}.$$
In the special case when $(U_i)=(0,0,0,1)$, from (2.3) it follows
that $\vec{w}=(0,0,0)$. If $(U_i)\neq (0,0,0,1)$, then $\vec{w}$ can
be nonzero, analogously as for frame dragging.

Now, let us consider the general formula for the parallel transport
of the considered frame $A_{i\alpha}$ in direction of the 4-vector
of velocity $V_i$. We introduce the tensor $P=P(U,V)$ given by
$$P_{ij}=\delta _{ij} - {\frac{1}{1 + U_{s}V_{s}}}
(V_{i}V_{j}+V_{i}U_{j}+U_{i}V_{j}+U_{i}U_{j}) +
2U_{j}V_{i},\eqno{(2.18)}$$ and accept {\em axiomatically} the
following relationship between the tensor $\phi_{ij}$ and the tensor
given by (2.16)
$$\frac{dA_{i\alpha}}{ds}A_{j\alpha}=
P_{ri}\phi_{rk}P_{kj},\eqno{(2.19)}$$ or in matrix form
$\frac{dA}{ds}A^T=P^T\phi P.$ Notice that both sides of (2.19) are
skew-symmetric matrices.

The tensor $P_{ij}$ is an orthogonal matrix. It can be verified by
using the identities $U_{i}U_{i}=V_{i}V_{i}=1$. Moreover, it has the
following property $P(U,V)=P(V,U)^{-1}$. Some other properties of
this tensor are given in \cite{CTbug} and a justification for its
appearance in (2.19) is given in \cite{TCbuc}. For example, it is
shown that using the standard addition, one can not uniquely
determine a 4-vector in the Minkowskian space-time which would
represent a relative 4-velocity of a point $B$ with respect to a
point $A$, assuming that $B$ moves with 4-velocity $V$ and $A$ moves
with 4-velocity $U$. So, the tensor $P(U,V)$ provides a transition
between velocities, i.e. $P_{ij}U_j=V_j$. The tensor $P$ with some
of its properties was independently found also by other authors
\cite{Fahn,Mat}.

In the special case $(U_{i})=(0,0,0,1)$, the tensor $P(U,V)$ is
given by
$$P = \left [\matrix{
1-{\frac{1}{\nu }}V_{1}^{2} &  -{\frac{1}{\nu }}V_{1}V_{2} &
-{\frac{1}{\nu }}V_{1}V_{3} &  V_{1}\cr & & & \cr
-{\frac{1}{\nu}}V_{2}V_{1} & 1-{\frac{1}{\nu }}V_{2}^{2} &
-{\frac{1}{\nu }}V_{2}V_{3} & V_{2}\cr & & & \cr -{\frac{1}{\nu}}
V_{3}V_{1} & -{\frac{1}{\nu }}V_{3}V_{2} & 1-{\frac{1}{\nu
}}V_{3}^{2} & V_{3}\cr & & & \cr -V_{1} & -V_{2} & -V_{3} & V_{4}\cr
}\right ], \eqno{(2.20)}$$ where $V_{1}, V_{2}, V_{3}, V_{4}$ are
given by (2.2), $\nu =1+V_{4}$, and this represents just a Lorentz
transformation (as a boost, without spatial rotation). Multiplying the
equation (2.19) by $A_{j\beta}$ and sum for $j$ we get
$$\frac{dA_{i\beta}}{ds}=P_{ri}\phi_{rk}P_{kj}A_{j\beta},\eqno{(2.21)}$$
and hence for the parallel displacement of an arbitrary vector $A_i$
we get
$$\frac{dA_i}{ds}=P_{ri}\phi_{rk}P_{kj}A_j.\eqno{(2.22)}$$
Particularly, for $A_i=V_i$, we obtain the equations of motion
$$\frac{dV_i}{ds}=P_{ri}\phi_{rk}P_{kj}V_j.\eqno{(2.23)}$$
The last equation $(i=4)$ of (2.23) is a consequence of the first
three equations, because if we multiply (2.23) by $V_i$ and sum for
$i=1,2,3,4$ we obtain the identity 0=0. The same is true for (2.22)
also.

Notice that the vectors $U_i$ and $V_i$ are tangent vectors of
different curves, parameterized for example via the time parameters.
Thus in (2.23) and the previous formulae, the 4-vector $U_i$ should
be taken at the point $(x',y',z',ict')$, where $t$ and $t'$ are
related by (2.5), because we must take into account the time which
is needed for the gravitational interaction to reach the test body.
But, if we take the values of $U_i$ at the same time $t$ as the
4-vector $V_i$, then the acceleration of the test body would be
changed of order $c^{-4}$, and so we will do that in this paper.

In the special case when $(U_i)=(0,0,0,1)$, the nonlinear connection
given by (2.22) and (2.23) is approximated \cite{T1} by a linear but
not metric connection, using Christoffel symbols $\Gamma^i_{jk}$,
such that $\Gamma^i_{jk}=-\Gamma^j_{ik}$. The Christoffel symbols
depend on the components of the tensor $\phi$ and it is verified
that the Einstein equations are satisfied up to $c^{-2}$ for such a
connection.

Two characteristics are essential for these equations of motion in
flat space:
{\it They are Lorentz invariant and they do
not use any special coordinate system}. But the inertial and
gravitational masses are different.

Let us consider the case of only one gravitational body in rest.
Then the tensor $\phi_{ij}$ and the equations of motion are
invariant under the transformation $\mu \rightarrow C\mu$, where $C$
is a constant. Thus the tensor $\phi$ and the equations of motion
are invariant under the gauge transformation $\ln \mu \rightarrow C
+ \ln\mu$, which is analogous to the Newtonian gauge transformation
$V\rightarrow V+C$, and analogous to the invariance of the equations
of motion in metric theories with respect to the transformation
$g_{ij}\rightarrow C\cdot g_{ij}$.

\vspace{-0.2cm}
\section{Geodesics Applied to Planetary Orbits, Light
Ray Trajectories and Gyroscope Precession}\label{sec:3}

Our coordinate origin will be chosen to be at the center of the Sun,
$U_1=U_2=U_3=0$ and the mass of each planet is assumed to be
negligible with respect to the mass of the Sun.

A straight calculation of the matrix $S=P^T\phi P$, where $\phi$ is
given by (2.17) and $P$ is given by (2.20), leads to
$$S_{41}=-S_{14}=i{\frac{\omega _z}{c}}V_2-i{\frac{\omega _y}{c}}V_3+
{\frac{a_x}{c{}^2}} \left( V_4+{\frac{(V_1){}^2}{1+V_4} }\right)
+{\frac{a_y}{c{}^2}} {\frac{V_1V_2}{1+V_4}}+{\frac{a_z}{c{}^2}
}{\frac{V_1V_3}{1+V_4}},$$
$$S_{42}=-S_{24}=i{\frac{\omega _x}{c}}V_3-i{\frac{\omega _z}{c}}V_1+
{\frac{a_x}{c{}^2}} {\frac{V_1V_2}{1+V_4}} +{\frac{a_y}{c{}^2}
}\left( V_4+{\frac{(V_2){}^2}{1+V_4}}\right) +{\frac{a_z}{c{}^2} }
{\frac{V_2V_3}{1+V_4}},$$
$$S_{43}=-S_{34}=i{\frac{\omega _y}{c}}V_1-i{\frac{\omega _x}{c}}V_2+
{\frac{a_x}{c{}^2}} {\frac{V_1V_3}{1+V_4}}+{\frac{a_y}{c{}^2}
}{\frac{V_2V_3}{1+V_4}} +{\frac{a_z}{c{}^2}}\left(
V_4+{\frac{(V_3)^2}{1+V_4}}\right),$$
$$S_{32}=-S_{23}={\frac{a_z}{c{}^2}}V_2-{\frac{a_y}{c{}^2}}V_3+
i{\frac{\omega _x}{c}}\left( V_4+{\frac{(V_1){}^2}{1+V_4} }\right)
+i{\frac{\omega_y}{c}}{\frac{V_1V_2}{1+V_4}}+i{\frac{\omega_z}{c}}
{\frac{V_1V_3}{1+V_4}},$$
$$S_{13}=-S_{31}={\frac{a_x}{c{}^2}}V_3-{\frac{a_z}{c{}^2}}V_1+
i{\frac{\omega _x}{c}} {\frac{V_1V_2}{1+V_4}} +i{\frac{\omega
_y}{c}}\left( V_4+{\frac{(V_2){}^2}{1+V_4}}\right) +i{\frac{\omega
_z}{c}} {\frac{V_2V_3}{1+V_4}},$$
$$S_{21}=-S_{12}={\frac{a_y}{c{}^2}}V_1-{\frac{a_x}{c{}^2}}V_2+
i{\frac{\omega _x}{c}} {\frac{V_1V_3}{1+V_4}}+i{\frac{\omega
_y}{c}}{\frac{V_2V_3}{1+V_4}} +i{\frac{\omega _z}{c}}\left(
V_4+{\frac{(V_3){}^2}{1+V_4}}\right) ,$$
$$S_{11}=S_{22}=S_{33}=S_{44}=0.\eqno{(3.1)}$$

Now by using equalities (3.1), the equations (2.23) become
$${\frac{dv_x}{dt}}=
\Bigl [(2-\beta ^{-2})a_x - {\frac{v_x}{c^2}
}(a_iv_i)\cdot(2+{\frac{1}{\beta (\beta +1)}})+2(v_yw_z-v_zw_y)\Bigr
],\eqno{(3.2a)}$$
$${\frac{dv_y}{dt}}=
\Bigl [(2-\beta ^{-2})a_y - {\frac{v_y}{c^2}
}(a_iv_i)\cdot(2+{\frac{1}{\beta (\beta +1)}})+2(v_zw_x-v_xw_z)\Bigr
],\eqno{(3.2b)}$$
$${\frac{dv_z}{dt}}=
\Bigl [(2-\beta ^{-2})a_z - {\frac{v_z}{c^2}
}(a_iv_i)\cdot(2+{\frac{1}{\beta (\beta +1)}})+2(v_xw_y-v_yw_x)\Bigr
],\eqno{(3.2c)}$$
$${\frac{d}{dt}}{\frac{1}{\sqrt{1-{\frac{v^2}{c^2}}}}} = {\frac{1}{c^2}}
(a_iv_i),\eqno{(3.2d)} $$ where $\beta = \Bigl (1-{\frac{v^2}{c^2}
}\Bigr )^{-1/2}$ and $a_iv_i=a_xv_x+a_yv_y+a_zv_z$. Indeed, (3.2d)
is a direct consequence of (2.23) for $i=4$, and then this equality
multiplied by $\displaystyle -\frac{v_x}{ic}$, $\displaystyle
-\frac{v_y}{ic}$, and $\displaystyle -\frac{v_z}{ic}$ should be
added to equations (2.23) for $i=1,2,3$, respectively in order to
find $dv_x/ds$, $dv_y/ds$, and $dv_z/ds$. It is easy to verify that
{\em if $U=(0,0,0,1)$, then the equation (3.2d) does not depend on
the matrix transformation $P$ applied to $\phi$, i.e. (3.2d) remains
unchanged if we take $\phi$ instead of $P^T\phi P$ in (2.23)}.

We will apply these equations in our special case. Using that $\mu
=1+{\frac{G M}{rc^2}}$, where $M$ is the mass of the Sun, from (2.3)
and (2.13) we obtain
$$(\phi_{ij})=\left [\matrix{
0 & 0 & 0 &{\frac{G M}{\mu r^3c^2}}x\cr & & & \cr 0 & 0 & 0
&{\frac{G M}{\mu r^3c^2}}y\cr & & & \cr 0 & 0 & 0 &{\frac{G M}{\mu
r^3c^2}}z\cr & & & \cr -{\frac{G M}{\mu r^3c^2}}x& -{\frac{G M}{\mu
r^3c^2}}y& -{\frac{G M}{\mu r^3c^2}}z&0\cr}\right ] ,\eqno{(3.3)}$$
and from the equations of motion (2.23), where the vector $V_i$ is
given by (2.2), can be found the components
$\displaystyle\frac{d^2x}{dt^2}=\frac{dv_x}{dt}$,
$\displaystyle\frac{d^2y}{dt^2}=\frac{dv_y}{dt}$, and
$\displaystyle\frac{d^2z}{dt^2}=\frac{dv_z}{dt}$. We replace
$v_z=0$, assuming that the test body moves in the $xy$-plane, and
thus, the equation for $i=3$ will be omitted. In this case, without
any approximation, the equations (3.2a), (3.2b), and (3.2d) reduce
to
$${\frac{d^2x}{dt^2}}={\frac{G M}{\mu r^3}}
\Bigl [(\beta ^{-2}-2)x + {\frac{v_x}{c^2}}(xv_x+yv_y)
(2+{\frac{1}{\beta (\beta +1)}})\Bigr ],\eqno{(3.4a)}$$
$${\frac{d^2y}{dt^2}}={\frac{G M}{\mu r^3}}
\Bigl [(\beta ^{-2}-2)y + {\frac{v_y}{c^2}}(xv_x+yv_y)
(2+{\frac{1}{\beta (\beta +1)}})\Bigr ].\eqno{(3.4b)}$$
$$\beta - \ln \Bigl (1+\frac{GM}{rc^2}\Bigr )=const.,\eqno{(3.4c)}$$
where (3.4c) is a solution of the differential equation (3.2d). This
equation can be written in the following form
$$U_iV_i-\ln \Bigl (1+\frac{GM}{rc^2}\Bigr )=const.,\eqno{(3.4d)}$$
where $U_i$ is the 4-vector of velocity of the Sun and $V_i$ is the
4-vector of velocity of the considered planet neglecting its mass.
The scalars $U_iV_i$ and the 3-dimensional distance $r$ determined
in the system where the Sun rests, are invariant of the choice of
the inertial coordinate system. Thus, the equation (3.4d) is Lorentz
invariant scalar equation. Indeed, the left side of (3.4d) is
proportional with the Hamiltonian, or more precisely the Hamiltonian is
given by
$${\cal H}=mc^2\Bigl( U_iV_i-\ln \Bigl (1+\frac{GM}{rc^2}\Bigr )\Bigr), 
\eqno{(3.4e)}$$
where the mass $m$ of the test body is negligible with respect to the
gravitational mass $M$.
According to the previous discussion, it does not
depend on the matrix transformation $P$. {\em Thus, $P$ does not
influence the energy of the moving body, but it influences only the
angular momentum of the moving body}. The previous discussion will
continue in 7.4, where the Lagrangian will be given.

Using that
$${\frac{d\varphi }{dt}}={\frac{d}{dt}}\arctan {\frac{y}{x}}=
{\frac{v_yx-v_xy}{r^2}}$$ for any angle $\varphi$, from (3.4a) and
(3.4b), we obtain
$${\frac{d}{dt}}\Bigl (r^2{\frac{d\varphi }{dt}}\Bigr ) =
{\frac{d^2y}{dt^2}}x-{\frac{d^2x}{dt^2}}y = {\frac{GM}{\mu
r^3c^2}}(v_yx-v_xy) (xv_x+yv_y)\Bigl (2+{\frac{1}{\beta (\beta
+1)}}\Bigr ) $$
$$=-r^2{\frac{d\varphi }{dt}}{\frac{G M}{\mu c^2}}\Bigl (
2+{\frac{1}{\beta (\beta +1)}}\Bigr ){\frac{d}{dt}}\Bigl
({\frac{1}{r}}\Bigr ).$$ Two cases will be considered.

\vspace{-0.2cm}
\subsection{Perihelion shift}

Assume that $v<<c$, and consider the planetary orbits. Then
$2+{\displaystyle \frac{1}{\beta (\beta +1)}} \approx 2.5$ so
neglecting the expressions of order $c^{-4}$ we can switch to
$${\frac{d}{dt}}\Bigl (r^2{\frac{d\varphi}{dt} }\Bigr ) = -{\frac{5}{2}}
{\frac{GM}{c^2}}\Bigl (r^2{\frac{d\varphi}{dt} }\Bigr )
{\frac{d}{dt}}\Bigl ({\frac{1}{r}}\Bigr ).$$ The solution of the
previous equation is
$$r^2{\frac{d\varphi}{dt}}=
C_2 \exp \Bigl ({\frac{-5}{2}}{\frac{G M}{rc^2}}\Bigr ), \quad
C_2=const.\eqno{(3.5)}$$

Further, using the metric $(dr)^2+r^2(d\varphi )^2-c^2t^2=ds^2$ in the flat space of Minkowski,
we obtain
$$\Bigl ({\frac{dr}{dt}}\Bigr )^2 +
\Bigl ({\frac{rd\varphi }{dt}}\Bigr )^2 = v^2, \qquad \Bigl (r^{-2}
{\frac{dr}{d\varphi }}\Bigr )^2 + r^{-2} = v^2 \Bigl
(r^2{\frac{d\varphi}{dt} }\Bigr )^{-2},$$ and $\rho =r^{-1}$
satisfies the equation
$$\Bigl ({\frac{d\rho }{d\varphi }}\Bigr )^2+\rho^2 = v^2C_2^{-2}
\exp \Bigl ({\frac{5G M\rho}{c^2}}\Bigr ).\eqno{(3.6)}$$

We are going to find $v^2$ from (3.4a) and (3.4b). By adding the
equation (3.4a) multiplied by $2v_x=2dx/dt$ and the equation (3.4b)
multiplied by $2v_y=2dy/dt$ and using that $xv_x+yv_y=rdr/dt$, we
get
$$\frac{dv^2}{dt}=\frac{GM}{\mu r^3}
\Bigl [(\beta^{-2}-2)2(xv_x+yv_y)+
2\frac{v^2}{c^2}(xv_x+yv_y)\cdot\frac{5}{2}\Bigr ]=$$
$$=\frac{GM}{\mu r^2}\frac{dr}{dt}
\Bigl (-2+3\frac{v^2}{c^2}\Bigr ) =\frac{2GM}{\mu }\frac{d\rho}{dt}
\Bigl (1-\frac{3}{2}\frac{v^2}{c^2}\Bigr ).$$ So, we obtain the
following differential equation
$$\Bigl (1-\frac{3}{2}\frac{v^2}{c^2}\Bigr )^{-1}
\frac{dv^2}{dt}=\frac{2GM}{\mu }\frac{d\rho}{dt}.$$ Replacing
$1/\mu$ with $1-\frac{GM\rho }{c^2}$ in the previous differential
equation and after some transformations, it becomes
$$\frac{-2}{3}c^2\frac{d\ln \Bigl (1-\frac{3}{2}\frac{v^2}{c^2}\Bigr )}
{dt}=2GM\frac{d}{dt}\Bigl (\rho -\frac{GM\rho^2}{2c^2}+C\Bigr ).$$
The solution by $v^2$ is given by
$$v^2=2GM\Bigl (\rho -\frac{GM\rho^2}{2c^2}
+C\Bigr )-3\frac{G^2M^2(\rho +C)^2}{c^2}.$$ After replacing this
value in (3.6) we obtain
$$\Bigl ({\frac{d\rho}{d\varphi}}\Bigr )^2+\rho^2=A+B\rho +
{\frac{6G ^2M^2}{c^2C_2^2}} \rho^2.\eqno{(3.7)}$$

Using that $C_2=\sqrt{G Ma(1-\epsilon^2)}$, where $a$ is the
semi-major axis and $\epsilon$ is the eccentricity, standard
calculations for the perihelion shift per orbit leads to the known
result
$$\Delta \varphi = {\frac{6GM\pi}{ac^2(1-\epsilon^2)}}.\eqno{(3.8)}$$

\vspace{-0.2cm}
\subsection{Deflection of the light rays near the Sun}

Let us consider the trajectory of a light ray near the Sun. We
denote by $R$ the radius of the Sun. In this case $\beta \rightarrow
\infty$, so
$${\frac{d}{dt}}\Bigl (r^2{\frac{d\varphi}{dt}}\Bigr ) = -2{\frac{G M}{c^2}}
\Bigl (r^2{\frac{d\varphi}{dt}}\Bigr ){\frac{d}{dt}}\Bigl
({\frac{1}{r}}\Bigr )$$ and its solution is
$$r^2{\frac{d\varphi}{dt}}=C_2\exp \Bigl (-2{\frac{G M}{rc^2}}\Bigr ),
\quad C_2=const.\eqno{(3.9)}$$ Analogously to (3.6) we obtain
$$\Bigl ({\frac{d\rho }{d\varphi}}\Bigr )^2 + \rho^2=v^2C_2^{-2}
\exp \Bigl ({\frac{4G M\rho}{c^2}}\Bigr )$$ and replacing $v=c$, we
get
$$\Bigl ({\frac{d\rho }{d\varphi}}\Bigr )^2 + \rho^2=c^2C_2^{-2}
\exp \Bigl ({\frac{4G M\rho}{c^2}}\Bigr ).\eqno{(3.10)}$$ The last
step was possible because it is easy to verify that the light has a
constant velocity $c$ in a gravitational field in orthonormal
coordinates.

If $r=R$, then $R{\frac{d\varphi}{dt} }=c$ and from (3.9) we get
$C_2=Rc\exp ({\frac{2G M}{Rc^2}})$. By replacing this value of $C_2$
into (3.10) we get
$$\Bigl ({\frac{d\rho }{d\varphi}}\Bigr )^2 + \rho^2={\frac{1}{R^2}}
\exp \Bigl ({\frac{4G M}{c^2}}\Bigl (\rho -{\frac{1}{R}}\Bigr )\Bigr
),
$$
$$\Bigl ({\frac{d\rho}{d\varphi}}\Bigr )^2 + \rho^2={\frac{1}{R^2}}
-{\frac{4G M}{R^3c^2}}+{\frac{4G M}{R^2c^2}}\rho .\eqno{(3.11)}$$
From (3.11) $\varphi $ can be determined as a function of $\rho$:
$$\varphi = \arccos {\frac{\rho R^2c^2-2G M}{Rc^2-2G M}},\eqno{(3.12)}$$
such that $\varphi =0$ if $\rho ={\frac{1}{R}}$. It is easy to
conclude from (3.12) that the angle of deflection of a light ray
near the Sun is equal to ${\frac{4G M}{Rc^2}}$. In \cite{T1} is
given a different proof for this angle.

\vspace{-0.2cm}
\subsection{Geodetic precession and the frame dragging effect}

Now we will deduce the formula for geodetic precession, simplifying
that the gravitational body rests in the chosen coordinate system
and hence also $\vec{w}=(0,0,0)$. We parallel transport the frame
$A_{i\alpha}$ from subsection 2.3, and assume that at the initial
moment it is given by the matrix (2.20). Then we calculate the
components $S_{3j}A_{j2}-S_{2j}A_{j3}$, $S_{1j}A_{j3}-S_{3j}A_{j1}$,
$S_{2j}A_{j1}-S_{1j}A_{j2}$, where the matrix $S$ is the same matrix
given by (3.1). Straight calculation of these components yields
$$S_{3j}A_{j2}-S_{2j}A_{j3} = 3i{\frac{a_yv_z-a_zv_y}{c^3}}, \;
S_{1j}A_{j3}-S_{3j}A_{j1} = 3i{\frac{a_zv_x-a_xv_z}{c^3}},$$
$$S_{2j}A_{j1}-S_{1j}A_{j2} = 3i{\frac{a_xv_y-a_yv_x}{c^3}}.$$
So, according to (2.21) we find
$$\frac{d(A_{32}-A_{23})/2}{ds} =
3i{\frac{a_yv_z-a_zv_y}{2c^3}},\; \frac{d(A_{13}-A_{31})/2}{ds} =
3i{\frac{a_zv_x-a_xv_z}{2c^3}},$$
$$\frac{d(A_{21}-A_{12})/2}{ds} =
3i{\frac{a_xv_y-a_yv_x}{2c^3}}. \eqno{(3.13)}$$

Having the transported matrix $A$, the following vector
${\frac{1}{2}}(A_{32}-A_{23},A_{13}-A_{31},A_{21}-A_{12})$,
represents just the 3-vector of the small spatial rotation. 
Hence for the required angular velocity we obtain the known GR formula
$$\vec{\Omega} = {\frac{3}{2}}{\frac{\vec{v}\times \vec{a}}{c^2}},\eqno{(3.14)}$$
which is confirmed to about 0.7\% using Lunar laser ranging data
\cite{D,WND}, and the recent GPB experiment.

In the previous phenomena the source of gravitation was in rest, but
for frame dragging effect it is necessary to consider a moving
source. In the system where the source rests the tensor $\phi$ is
well known and hence it is well known in any other system. Similar
calculations to the previous yield the same formula for frame
dragging \cite{TC1} as the GR formula \cite{W}.

\vspace{-0.2cm}
\section{Periastron Shift of the Binary Systems}\label{sec:5}

Let us consider an arbitrary binary system, for example a pulsar and
its companion. In this section the periastron shift of the binary
system will be calculated, assuming that both bodies are moving in
the $xy$-plane. Let $m$ be the mass of a pulsar and $M$ be the mass
of its companion, and let us choose the coordinate system such that
at the initial moment $\vec{r}_b=(0,0,0)$ and
$\frac{\vec{r}_b}{dt}=(0,0,0)$, where $\vec{r}_b$ is the barycenter
(2.15) of the two bodies. We denote by $(x,y,0)$ the coordinates of
the pulsar, and by $(x',y',0)$ the coordinates of its companion.

Let the 4-vectors of velocity of the pulsar and its companion are
given by (2.2) and (2.1) respectively, where $u_z=v_z=0$. It is
convenient to use the notations
$$R=\sqrt{(x-x')^2+(y-y')^2},\enskip r=\sqrt{x^2+y^2}, \enskip \rho =1/r,
\enskip\ r\approx \frac{M}{M+m}R.$$ If we make the replacements
$\frac{x-x'}{R}=\cos \alpha$ and $\frac{y-y'}{R}=\sin \alpha$, then
$$\cos \alpha = \frac{x}{r},\quad \sin \alpha =
\frac{y}{r},\quad x'/y'=x/y,\quad u_x\approx -v_x\frac{m}{M}, \quad
u_y\approx -v_y\frac{m}{M}.$$ The acceleration of the pulsar (2.23),
at the initial moment with the initial conditions, can be simplified
into the following form
$$\frac{d^2x}{dt^2}=-\Bigl (1-\frac{v^2}{2c^2}\Bigr )c^2S_{14}-
\frac{v_x}{c^2}(a_xv_x+a_yv_y)+icv_yS_{12},\eqno{(4.1a)}$$
$$\frac{d^2y}{dt^2}=-\Bigl (1-\frac{v^2}{2c^2}\Bigr )c^2S_{24}-
\frac{v_y}{c^2}(a_xv_x+a_yv_y)-icv_xS_{12},\eqno{(4.1b)}$$ 
where the components of the matrix $S=P(U,V)^T\phi P(U,V)$ should be
calculated analogously to (3.1). In order to avoid large expressions
for $c^{-2}$ approximation, it is sufficient to use the components
(3.1), replacing $v_x$ by $v_x-u_x$ and $v_y$ by $v_y-u_y$. Hence
for $S_{14}$, $S_{24}$, and $S_{12}$ we obtain
$$S_{14}=\Bigl [-a_x\Bigl (\frac{1}{\sqrt{1-
\frac{(\vec{v}-\vec{u})^2}{c^2}}}-\frac{(v_x-u_x)^2}{2c^2}\Bigr)+$$
$$+a_y\frac{(v_x-u_x)(v_y-u_y)}{2c^2}-w_z(v_y-u_y)\Bigr
]\frac{1}{c^2},$$
$$S_{24}=\Bigl [-a_y\Bigl (
\frac{1}{\sqrt{1-\frac{(\vec{v}-\vec{u})^2}{c^2}}}-
\frac{(v_y-u_y)^2}{2c^2}\Bigr)+$$
$$+a_x\frac{(v_x-u_x)(v_y-u_y)}{2c^2}+w_z(v_x-u_x)\Bigr
]\frac{1}{c^2},$$
$$S_{12}=-\frac{i}{c}\Bigl [\frac{a_x}{c^2}(v_y-u_y)-
\frac{a_y}{c^2}(v_x-u_x)+w_z\Bigr ] .$$

According to (2.6) up to $c^{-2}$, we have
$$\frac{1-\frac{u^2}{c^2}}{(1-\frac{u^2}{c^2}\sin^2\theta)^{3/2}}
=\frac{1-\frac{u^2}{c^2}}
{(1-\frac{u^2}{c^2})^{3/2}(1+\frac{u^2}{c^2}\cos^2\theta)^{3/2}}=$$
$$
=\frac{1}{\sqrt{1-\frac{u^2}{c^2}}}\frac{1}{(1+\frac{1}{c^2}
\frac{m^2}{M^2}(\vec{r}'\cdot \frac{\vec{r}}{r})^2)^{3/2}}=
\frac{1}{\sqrt{1-\frac{u^2}{c^2}}}\frac{1}{(1+\frac{1}{c^2}
\frac{m^2}{(M+m)^2}(\frac{dR}{dt})^2)^{3/2}}.$$ Using the results
from 2.1 and 2.3, the components $a_x$, $a_y$, $w_z$ are given by
$$a_x=-\frac{x}{r}\frac{1}{\sqrt{1-\frac{u^2}{c^2}}}\frac{GM}{R^2}
\Bigl (1-\frac{G(M+2m)}{Rc^2}\Bigr )\lambda^{-3},$$
$$a_y=-\frac{y}{r}\frac{1}{\sqrt{1-\frac{u^2}{c^2}}}\frac{GM}{R^2}
\Bigl (1-\frac{G(M+2m)}{Rc^2}\Bigr )\lambda^{-3},$$
$$w_z=\frac{Gm}{R^2c^2}\frac{v_xy-v_yx}{r}\lambda^{-3},\qquad
\lambda
=\sqrt{1+\frac{1}{c^2}\frac{m^2}{(M+m)^2}\Bigl(\frac{dR}{dt}\Bigr)^2}.$$

Using the equalities between $v_x$, $u_x$; $v_y$, $u_y$; $r$, $R$
and so on, (4.1) can be reduced to the following form
$$\frac{d^2\vec{r}}{dt^2}=-\frac{\vec{R}}{R}\frac{GM}{R^2}\Bigl [
1+\frac{V^2}{c^2}\frac{M^2+4Mm+2m^2}{(M+m)^2}-
\frac{G(M+2m)}{Rc^2}-$$
$$-\frac{3}{2c^2}\frac{m^2}{(M+m)^2}\Bigl
(\frac{dR}{dt}\Bigr )^2\Bigr ]+
\vec{V}\frac{dR}{dt}\frac{GM}{R^2}\Bigl (\frac{M}{M+m}+
\frac{3}{2}\Bigr )\frac{1}{c^2},\eqno{(4.2)}$$ where $\vec{V}$ is
the relative velocity of the pulsar with respect to its companion.

Analogously to this acceleration, the acceleration of the body with
mass $M$ (pulsar companion) at the initial moment is given by
$$\frac{d^2\vec{r}'}{dt^2}=\frac{\vec{R}}{R}\frac{Gm}{R^2}\Bigl [
1+\frac{V^2}{c^2}\frac{m^2+4Mm+2M^2}{(M+m)^2} -\frac{G(2M+m)}{Rc^2}-
$$
$$-\frac{3}{2c^2}\frac{M^2}{(M+m)^2}\Bigl
(\frac{dR}{dt}\Bigr )^2\Bigr ] -\vec{V}\frac{dR}{dt}\frac{Gm}{R^2}
\Bigl (\frac{m}{M+m}+\frac{3}{2}\Bigr )\frac{1}{c^2}.\eqno{(4.3)}$$
Subtracting the equation (4.3) from (4.2), after some
transformations we get
$$\frac{d^2\vec{R}}{dt^2}=-\frac{\vec{R}}{R}\frac{G(M+m)}{R^2}\Bigl [
1+\frac{V^2}{c^2}\frac{M^2+5Mm+m^2}{(M+m)^2}-
\frac{G(M^2+4Mm+m^2)}{Rc^2(M+m)}$$
$$-\frac{3}{2c^2}\frac{Mm}{(M+m)^2}\Bigl (\frac{dR}{dt}
\Bigr )^2\Bigr ]+
\vec{V}\frac{dR}{dt}\frac{G}{R^2}\frac{5M^2+6Mm+5m^2}{2(M+m)c^2}.
\eqno{(4.4)}$$

All variables in (4.4) are related to the relative motion and so the
assumption about the initial moment has no role. Now, having the
system of equations (4.4) for the relative motion of a body with
mass $m$ with respect to the body with mass $M$, we can calculate
the periastron shift in two steps, analogously as it has been made
for the perihelion shift in section \ref{sec:3}. The first step
consists of finding an equation analogous to (3.6) in the same way
as in section \ref{sec:3}. In the second step we sum the first
equation in (4.4) multiplied by $2V_x$ and the second equation of
(4.4) multiplied by $2V_y$. That equation can be integrated and the
value of $V^2$ can be found. After these two steps the periastron
shift can be obtained. We present only the final results of these
two steps avoiding the long algebraic and differential calculations.

The first step from the system (4.4) yields the following equation
$$\Bigl (\frac{d\frac{1}{R}}{d\varphi}\Bigr )^2+\frac{1}{R^2}=V^2C_2^{-2}
\Bigl [1+\frac{5M^2+6Mm+5m^2}{M+m}\frac{G}{Rc^2}\Bigr
].\eqno{(4.5)}$$

The second step is more complicated. Although
$\frac{3}{2c^2}\frac{Mm}{(M+m)^2}\Bigl (\frac{dR}{dt}\Bigr )^2$ has
no influence in (4.5), it has a significant role in $V^2$, but we
will see that it has no role in the periastron shift. In order to
simplify the system (4.4), one can prove that
$\frac{3}{2c^2}\frac{Mm}{(M+m)^2}\Bigl (\frac{dR}{dt}\Bigr )^2$ has
no influence on the periastron shift. The proof is standard and thus
we omit it.

The second step from the modified system (4.4) yields
$$V^2=2\frac{G(M+m)}{R}-\frac{4G^2}{R^2c^2}(M^2+m^2)
+ \frac{C}{Rc^2}+K,\eqno{(4.6)}$$ where $C$ and $K$ are mutually
dependent constants, which have no role in the periastron shift.
Now, analogously to (3.7) from (4.5) and (4.6), we get
$$\Bigl (\frac{d\frac{1}{R}}{d\varphi}\Bigr )^2+\frac{1}{R^2}
=A+B\frac{1}{R}+\frac{6G^2(M+m)^2}{C_2^2c^2}\frac{1}{R^2}.$$ Using
that $C_2^2=G(M+m)a_r(1-\epsilon ^2)$, similar to the calculations
for the perihelion shift in section \ref{sec:3}, for the periastron
shift we obtain
$$\Delta \varphi =\frac{6\pi G(M+m)}{a_r(1-\epsilon ^2)c^2},
\eqno{(4.7)} $$ where $a_r$ is the semi-major axis of the relative
orbit and $\epsilon$ is the eccentricity of the orbit. This result
is the same as in the GR.

\vspace{-0.2cm}
\section{Barycenter of Two Bodies}\label{sec:6}

We proceed with the problem of two bodies considering their
barycenter. We shall employ the same notations as in the previous
section, and we will use the same coordinate system with the
assumptions about the initial moment. Now, we prove that in the
chosen coordinate system the barycenter coincides with the
coordinate origin. According to (2.15) we have
$$\vec{r}=-\vec{r}'\frac{M(1+\frac{u^2}{2c^2}-\frac{Gm}{2Rc^2})}
{m(1+\frac{v^2}{2c^2}-\frac{GM}{2Rc^2})}+\vec{r}_b\frac
{M(1+\frac{u^2}{2c^2}-\frac{Gm}{2Rc^2})+m(1+\frac{v^2}{2c^2}-\frac{GM}{2Rc^2})}
{m(1+\frac{v^2}{2c^2}-\frac{GM}{2Rc^2})}.$$ Since
$$-\frac{M(1+\frac{u^2}{2c^2}-\frac{Gm}{2Rc^2})}
{m(1+\frac{v^2}{2c^2}-\frac{GM}{2Rc^2})}
=-\frac{M(1+\frac{V^2}{2c^2}\frac{m^2}{(M+m)^2}-\frac{Gm}{2Rc^2})}
{m(1+\frac{V^2}{2c^2}\frac{M^2}{(M+m)^2}-\frac{GM}{2Rc^2})}=$$
$$=-\frac{M}{m} \Bigl
(1-\frac{M-m}{2(M+m)}\frac{V^2}{c^2}+\frac{1}{2}
\frac{G(M-m)}{Rc^2}\Bigr ),$$ at each moment it is satisfied
$$\frac{1}{M}\vec{r}=-\frac{1}{m}\vec{r}'
\Bigl (1-\frac{M-m}{2(M+m)}\frac{V^2}{c^2}+
\frac{G(M-m)}{2Rc^2}\Bigr )+$$
$$+\vec{r}_b\Bigl [\frac{1}{M}+\frac{1}{m}\Bigl (
1-\frac{M-m}{2(M+m)}\frac{V^2}{c^2}+ \frac{G(M-m)}{2Rc^2}\Bigr
)\Bigr ]. \eqno{(5.1)}$$

The accelerations (4.2) and (4.3) are given with respect to the
coordinate system such that at the initial moment the coordinate
origin coincides with the barycenter $\vec{r}_b$ and
$d\vec{r}_b/dt=(0,0,0)$ at the initial moment. But, if we replace
$\vec{r}$ by $\vec{r}-\vec{r}_b$ in (4.2) and replace $\vec{r}'$ by
$\vec{r}'-\vec{r}_b$ in (4.3), then (4.2) and (4.3) will be true for
each $t$. Further, we will use these modified equations and will
prove that at each moment
$$\frac{1}{M}\frac{d^2\vec{r}}{dt^2}=-\frac{1}{m}\frac{d^2}{dt^2}
\Bigl [\vec{r}'\Bigl (1-\frac{M-m}{2(M+m)}\frac{V^2}{c^2}+
\frac{G(M-m)}{2Rc^2}\Bigr )\Bigr ]+ \Bigl
(\frac{1}{M}+\frac{1}{m}\Bigr )\frac{d^2\vec{r}_b}{dt^2},$$ i.e.
$$\frac{1}{M}\frac{d^2\vec{r}}{dt^2}+\frac{1}{m}\frac{d^2\vec{r}'}
{dt^2}=$$
$$= \frac{d^2}{dt^2}\Bigl [\frac{\vec{R}}{M+m}\Bigl (
-\frac{M-m}{2(M+m)}\frac{V^2}{c^2}+ \frac{G(M-m)}{2Rc^2}\Bigr )\Bigr
]+ \Bigl (\frac{1}{M}+\frac{1}{m}\Bigr
)\frac{d^2\vec{r}_b}{dt^2}.\eqno{(5.2)}$$

According to the modified equations (4.2) and (4.3), the left side
of (5.2) is equal to
$$\frac{1}{M}\frac{d^2\vec{r}}{dt^2}+\frac{1}{m}
\frac{d^2\vec{r}'}{dt^2}=$$
$$=\frac{1}{M}\Bigl [\frac{\vec{R}}{R}\frac{GM}{R^2}
\frac{V^2}{c^2}\frac{M^2}{(M+m)^2}
+\vec{V}\frac{dR}{dt}\frac{GM^2}{(M+m)R^2c^2}+$$ $$+
\frac{\vec{R}}{R}\frac{G^2Mm}{Rc^2}\frac{1}{R^2}
+\frac{3}{2}\frac{\vec{R}}{R}\frac{GM}{R^2c^2}
\frac{m^2}{(M+m)^2}\Bigl (\frac{dR}{dt}\Bigr )^2\Bigr ]-$$
$$-\frac{1}{m}\Bigl [\frac{\vec{R}}{R}\frac{Gm}{R^2}
\frac{V^2}{c^2}\frac{m^2}{(M+m)^2}
+\vec{V}\frac{dR}{dt}\frac{Gm^2}{(M+m)R^2c^2}+$$ $$+
\frac{\vec{R}}{R}\frac{G^2Mm}{Rc^2}\frac{1}{R^2}
+\frac{3}{2}\frac{\vec{R}}{R}\frac{Gm}{R^2c^2}
\frac{M^2}{(M+m)^2}\Bigl (\frac{dR}{dt}\Bigr )^2\Bigr ]+\Bigl
(\frac{1}{M}+\frac{1}{m}\Bigr )\frac{d^2\vec{r}_b}{dt^2}=$$
$$=\frac{\vec{R}}{R}\frac{G}{R^2}\frac{V^2}{c^2}\frac{M-m}{M+m}
+\vec{V}\frac{dR}{dt}\frac{G}{R^2c^2}\frac{M-m}{M+m}-$$ $$-
\frac{\vec{R}}{R}\frac{G^2(M-m)}{Rc^2}\frac{1}{R^2}
-\frac{3}{2}\frac{\vec{R}}{R}\frac{G}{R^2c^2} \frac{M-m}{M+m}\Bigl
(\frac{dR}{dt}\Bigr )^2 +\Bigl (\frac{1}{M}+\frac{1}{m}\Bigr
)\frac{d^2\vec{r}_b}{dt^2}.$$ Hence, it is sufficient to prove that
$$\frac{d^2}{dt^2}\Bigl [\frac{\vec{R}}{M+m}\Bigl (
-\frac{M-m}{2(M+m)}\frac{V^2}{c^2}+ \frac{G(M-m)}{2Rc^2}\Bigr )\Bigr
]=$$
$$=\frac{\vec{R}}{R}\frac{G}{R^2}\frac{V^2}{c^2}\frac{M-m}{M+m}
+\vec{V}\frac{dR}{dt}\frac{G}{R^2c^2}\frac{M-m}{M+m}-$$ $$-
\frac{\vec{R}}{R}\frac{G^2(M-m)}{Rc^2}\frac{1}{R^2}
-\frac{3}{2}\frac{\vec{R}}{R}\frac{G}{R^2c^2} \frac{M-m}{M+m}\Bigl
(\frac{dR}{dt}\Bigr )^2.$$ After multiplication with
$\frac{2c^2}{G}\frac{M+m}{M-m}$, we should prove that
$$\frac{d^2}{dt^2}\frac{\vec{R}}{R}-\frac{1}{G(M+m)}
\frac{d^2}{dt^2}(\vec{R}(\vec{V}\cdot \vec{V}))=
2\frac{\vec{R}}{R}\frac{1}{R^2}V^2+$$
$$+2\vec{V}\frac{dR}{dt}\frac{1}{R^2}
-2\frac{\vec{R}}{R}\frac{(M+m)G}{R^3}-3\frac{\vec{R}}{R^3} \Bigl
(\frac{dR}{dt}\Bigr )^2.\eqno{(5.3)}$$ Using the identities
$$\frac{d^2}{dt^2}\frac{\vec{R}}{R}= -2\vec{V}\frac{dR}{dt}\frac{1}{R^2}
-\frac{\vec{R}}{R^3}V^2+3\frac{\vec{R}}{R^3}\Bigl
(\frac{dR}{dt}\Bigr )^2\quad \hbox{and} \quad
\frac{dR}{dt}=\vec{V}\cdot \frac{\vec{R}}{R},$$ the identity (5.3)
is equivalent to
$$-\frac{1}{G(M+m)}\frac{d^2}{dt^2}(\vec{R}(\vec{V}\cdot
\vec{V}))=$$
$$=3\frac{\vec{R}}{R^3}V^2+4\vec{V}\frac{dR}{dt}\frac{1}{R^2}
-6\frac{\vec{R}}{R^3}\Bigl (\frac{dR}{dt}\Bigr )^2-
2\frac{\vec{R}}{R}\frac{(M+m)G}{R^3}.\eqno{(5.4)}$$ A straight
calculation and using the identity $V^2=\frac{2G(M+m)}{R}+const.$
one verifies the identity (5.4), i.e. (5.2).

From (5.1) and (5.2) it follows that
$$\Bigl (\frac{1}{M}+\frac{1}{m}\Bigr )\frac{d^2\vec{r}_b}{dt^2}=
\frac{d^2}{dt^2}\Bigl [\vec{r}_b\Bigl (\frac{1}{M}+\frac{1}{m} \Bigl
( 1-\frac{M-m}{2(M+m)}\frac{V^2}{c^2}+ \frac{G(M-m)}{2Rc^2}\Bigr
)\Bigr )\Bigr ],$$ i.e.
$$\frac{d^2}{dt^2}\Bigl [\vec{r}_b\Bigl (\frac{G(M+m)}{R}-V^2\Bigr )\Bigr ]=0.
\eqno{(5.5)}$$ Hence it follows
$$\vec{r}_b\Bigl (\frac{G(M+m)}{R}-V^2\Bigr )=A+Bt,$$
where $A$ and $B$ are constants, assuming that the initial moment is
$t=0$. Since $\vec{r}_b=(0,0,0)$ for $t=0$ by assumption, we obtain
$A=0$. By assumption we also have $d\vec{r}_b/dt=(0,0,0)$ for $t=0$,
and hence $B=0$. Thus, $\vec{r}_b \equiv (0,0,0)$, if we assume that
$\vec{r}_b=(0,0,0)$ and $d\vec{r}_b/dt=(0,0,0)$ at the initial
moment. As a consequence, the formulae (4.2) and (4.3) are true not
only at the initial moment, but along the whole trajectory.

\vspace{-0.2cm}
\section{Equations of Motion for $n$-Body Problem
and Their Relationship with the GR Equations}\label{sec:8}

Now, we will consider the $n$-body problem, i.e. the equations of
motion of $n$ bodies with arbitrary masses. Assume that all bodies
are compressed into points, and hence we neglect their angular
momenta. We will obtain the equations of motion in explicit form
using 3-vectors of distances between the bodies, and their
velocities and accelerations.

Let a system of $n$ bodies with masses $m_1$, $m_2$, $\cdots $,
$m_n$, with initial positions and initial velocities be given. We
denote by $\vec{r}_k$ and $\vec{v}_k$ the 3-radius vector and
3-vector of velocity of the body with mass $m_k$, and denote
$r_{ij}=\vert \vec{r}_i-\vec{r}_j\vert$ for $i\neq j$. We will write
the equations only for the motion of the $j$-th body under the
gravitation of the $i$-th body and then follows summation for all
$i\neq j$.

Analogously as in section \ref{sec:5}, the components of the matrix
$S$ are
$$S_{14}=\Bigl [-a_x\Bigl (1+\frac{(v_y-u_y)^2}{2c^2}+
\frac{(v_z-u_z)^2}{2c^2}\Bigr )+
a_y\frac{(v_x-u_x)(v_y-u_y)}{2c^2}+$$
$$+a_z\frac{(v_x-u_x)(v_z-u_z)}{2c^2}+
w_y(v_z-u_z)-w_z(v_y-u_y)\Bigr ]\frac{1}{c^2},$$
$$S_{24}=\Bigl [a_x\frac{(v_x-u_x)(v_y-u_y)}{2c^2}-
a_y\Bigl (1+\frac{(v_x-u_x)^2}{2c^2}+\frac{(v_z-u_z)^2} {2c^2}\Bigr
)+$$
$$+a_z\frac{(v_y-u_y)(v_z-u_z)}{2c^2}+
w_z(v_x-u_x)-w_x(v_z-u_z)\Bigr ]\frac{1}{c^2},$$
$$S_{34}=\Bigl [a_x\frac{(v_x-u_x)(v_z-u_z)}{2c^2}+a_y
\frac{(v_y-u_y)(v_z-u_z)}{2c^2}-$$
$$-a_z\Bigl (
1+\frac{(v_x-u_x)^2}{2c^2}+\frac{(v_y-u_y)^2}{2c^2}\Bigr )+
w_x(v_y-u_y)-w_y(v_x-u_x)\Bigr ]\frac{1}{c^2},$$
$$S_{12}=-\frac{i}{c}\Bigl [\frac{a_x}{c^2}(v_y-u_y)-
\frac{a_y}{c^2}(v_x-u_x)+w_z\Bigr ],$$
$$S_{23}=-\frac{i}{c}\Bigl [\frac{a_y}{c^2}(v_z-u_z)-
\frac{a_z}{c^2}(v_y-u_y)+w_x\Bigr ],$$
$$S_{31}=-\frac{i}{c}\Bigl [\frac{a_z}{c^2}(v_x-u_x)-
\frac{a_x}{c^2}(v_z-u_z)+w_y\Bigr ],$$ where
$(v_x,v_y,v_z)=\vec{v}_j$ and $(u_x,u_y,u_z)=\vec{v}_i$. Notice that
the orthogonal matrix $P$ from (2.18) besides the Lorentz
transformation of velocity $\vec{u}-\vec{v}$ contains also a spatial rotation 
determined by the 3-vector $(\vec{u}\times \vec{v})/c^2$.
This small spatial rotation will be taken into account by changing the
components of the tensor $\phi$, i.e. of $a_x,a_y,a_z$, while the
influence on $w_x,w_y,w_z$ is of order $c^{-4}$. The previous
equations can be written in the following compact form
$$\vec{S}=\Bigl [\vec{a}\Bigl (1+\frac{(\vec{v}_j-\vec{v}_i)^2}{2c^2}\Bigr )
-\frac{\vec{v}_j-\vec{v}_i}{2c^2}(\vec{a}\cdot
(\vec{v}_j-\vec{v}_i))+ (\vec{v}_j-\vec{v}_i)\times \vec{w}\Bigr
]\frac{1}{c^2},$$
$$\vec{S}^*=-\frac{i}{c^3}[\vec{a}\times
(\vec{v}_j-\vec{v}_i)]-\frac{i}{c}\vec{w},\eqno{(6.1)}$$ where
$\vec{S}=(S_{41},S_{42},S_{43})$ and
$\vec{S}^*=(S_{23},S_{31},S_{12})$.

Further, the components of the tensor $\phi$ caused by the body with
mass $m_i$ are given by
$$a_x=-(x-x')\frac{Gm_i}{R^3\lambda^3\mu}\Bigl (1+\frac{u^2}{2c^2}\Bigr )
-\frac{Gm_i}{r_{ij}^3c^2}\Bigl [(\vec{r}_j-\vec{r}_i)\times
((\vec{r}_j-\vec{r}_i)\times \dot{\vec{v}}_i)\Bigr ]_x-$$
$$-(y-y')\frac{v_xu_y-v_yu_x}{c^2}\frac{Gm_i}{R^3}+
(z-z')\frac{v_zu_x-v_xu_z}{c^2}\frac{Gm_i}{R^3},$$
$$a_y=-(y-y')\frac{Gm_i}{R^3\lambda^3\mu}\Bigl (1+\frac{u^2}{2c^2}\Bigr )
-\frac{Gm_i}{r_{ij}^3c^2}\Bigl [(\vec{r}_j-\vec{r}_i)\times
((\vec{r}_j-\vec{r}_i)\times \dot{\vec{v}}_i)\Bigr ]_y-$$
$$-(z-z')\frac{v_yu_z-v_zu_y}{c^2}\frac{Gm_i}{R^3}+
(x-x')\frac{v_xu_y-v_yu_x}{c^2}\frac{Gm_i}{R^3},$$
$$a_z=-(z-z')\frac{Gm_i}{R^3\lambda^3\mu}\Bigl (1+\frac{u^2}{2c^2}\Bigr )
-\frac{Gm_i}{r_{ij}^3c^2}\Bigl [(\vec{r}_j-\vec{r}_i)\times
((\vec{r}_j-\vec{r}_i)\times \dot{\vec{v}}_i)\Bigr ]_z-$$
$$-(x-x')\frac{v_zu_x-v_xu_z}{c^2}\frac{Gm_i}{R^3}+
(y-y')\frac{v_yu_z-v_zu_y}{c^2}\frac{Gm_i}{R^3},$$
$$w_x=\frac{Gm_i}{c^2R^3}[(y-y')u_z-u_y(z-z')],$$
$$w_y=\frac{Gm_i}{c^2R^3}[(z-z')u_x-u_z(x-x')],\qquad
w_z=\frac{Gm_i}{c^2R^3}[(x-x')u_y-u_x(y-y')],$$ where
$(x,y,z)=\vec{r}_j$, $(x',y',z')=\vec{r}_i$, $R=r_{ij}$, $\lambda
=1+\frac{1}{2c^2}\Bigl (\vec{u}\cdot
\frac{\vec{r}_j-\vec{r}_i}{r_{ij}}\Bigr )^{2}$,
$$\mu =\Bigl (1+\frac{Gm_i}{r_{ij}c^2}\Bigr )
\Bigl (1+\frac{2Gm_j}{r_{ij}c^2}\Bigr )\prod _{k\neq i,j} \Bigl
[\Bigl (1+\frac{Gm_k}{r_{ki}c^2}\Bigr ) \Bigl
(1+\frac{Gm_k}{r_{kj}c^2}\Bigr )\Bigr ].$$ The last two terms in
$a_x$, $a_y$, and $a_z$ are added as influence from the spatial
rotation given by the vector $(\vec{u}\times \vec{v})/c^2$. These
equalities in vector form can be written as
$$\vec{a}=-\frac{(\vec{r}_j-\vec{r}_i)Gm_i}{r_{ij}^3}\Bigl [
1-\frac{3}{2}\frac{[\vec{v}_i\cdot (\vec{r}_j-\vec{r}_i)]^2}
{r_{ij}^2c^2}-$$ $$-\frac{G(m_i+2m_j)}{r_{ij}c^2}-\sum_{k\neq i,j}
\Bigl (\frac{Gm_k}{r_{ki}c^2}+\frac{Gm_k}{r_{kj}c^2}\Bigr )+
\frac{v_i^2}{2c^2}\Bigr ]-$$
$$-\frac{Gm_i}{r_{ij}^3c^2}(\vec{r}_j-\vec{r}_i)\times
((\vec{r}_j-\vec{r}_i)\times \dot{\vec{v}}_i)-
\frac{Gm_i}{r_{ij}^3c^2}(\vec{r}_j-\vec{r}_i)\times (\vec{v}_j\times
\vec{v}_i). \eqno{(6.2)}$$

For the equations of motion of the $j$-th body, analogously to (4.1)
we obtain
$$\frac{d^2x}{dt^2}=\Bigl (1-\frac{v^2}{2c^2}\Bigr )
\bigl [-(S_{41}v_x^2+S_{42}v_xv_y+ S_{43}v_xv_z)
+ic(S_{12}v_y+S_{13}v_z+icS_{14})\bigr ],$$
$$\frac{d^2y}{dt^2}=\Bigl (1-\frac{v^2}{2c^2}\Bigr )
\bigl [-(S_{41}v_xv_y+S_{42}v_y^2+ S_{43}v_yv_z)+
ic(S_{21}v_x+S_{23}v_z+icS_{24})\bigr ],$$
$$\frac{d^2z}{dt^2}=\Bigl (1-\frac{v^2}{2c^2}\Bigr )
\bigl [-(S_{41}v_xv_z+S_{42}v_yv_z+ S_{43}v_z^2)+
ic(S_{31}v_x+S_{32}v_y+icS_{34})\bigr ].$$ 
In vector form they can be written as
$$\frac{d^2\vec{r}_j}{dt^2}=-\vec{v}_j(\vec{S}\cdot \vec{v})
\Bigl (1-\frac{v^2}{2c^2}\Bigr ) + \vec{S}\Bigl (
1-\frac{v^2}{2c^2}\Bigr )c^2+ic\Bigl (1-\frac{v^2}{2c^2}\Bigr )
(\vec{v}\times \vec{S}^*).\eqno{(6.3)}$$ Finally, after summation
for all $i\neq j$ and after many transformations, the equations
(6.3) become
$$\frac{d^2\vec{r}_j}{dt^2}=\sum_{i\neq j}\Bigl \{
-\frac{(\vec{r}_j-\vec{r}_i)Gm_i}{r_{ij}^3}\Bigl [
1-\frac{3}{2}\frac{[\vec{v}_i\cdot (\vec{r}_j-\vec{r}_i)]^2}
{r_{ij}^2c^2}-$$
$$-\frac{G(m_i+2m_j)}{r_{ij}c^2}-\sum_{k\neq i,j}
\Bigl (\frac{Gm_k}{r_{ki}c^2}+\frac{Gm_k}{r_{kj}c^2}\Bigr
)+\frac{v_i^2}{c^2}-2\frac{\vec{v}_i\cdot \vec{v}_j}{c^2}+
\frac{(\vec{v}_i-\vec{v}_j)^2}{c^2}\Bigr ]-$$
$$-\frac{Gm_i}{r_{ij}^3c^2}(\vec{r}_j-\vec{r}_i)\times
((\vec{r}_j-\vec{r}_i)\times \dot{\vec{v}}_i)
+\frac{3}{2}\frac{Gm_i}{r_{ij}^3c^2}(\vec{v}_j-\vec{v}_i)
[(\vec{r}_j-\vec{r}_i)\cdot (\vec{v}_j-\vec{v}_i)]+$$
$$+
\frac{Gm_i}{r_{ij}^3c^2}(\vec{v}_j-\vec{v}_i)
[(\vec{r}_j-\vec{r}_i)\cdot \vec{v}_j]\Bigr \} .\eqno{(6.4)}$$

Now, let us distinguish the terms in (6.4) which are Lorentz
invariant. From (6.4) we obtain
$$\frac{d^2\vec{r}_j}{dt^2}=\sum_{i\neq j}\Bigl \{
-\frac{(\vec{r}_j-\vec{r}_i)Gm_i}{r_{ij}^3}\Bigl [
1-\frac{3}{2}\frac{[\vec{v}_i\cdot
(\vec{r}_j-\vec{r}_i)]^2}{r_{ij}^2c^2} +
\frac{v_i^2}{c^2}-2\frac{\vec{v}_i\cdot \vec{v}_j}{c^2}\Bigr ]+$$
$$+\frac{Gm_i}{r_{ij}^3c^2}(\vec{v}_j-\vec{v}_i)
[(\vec{r}_j-\vec{r}_i)\cdot \vec{v}_j]\Bigr \} + \hbox{Lorentz}\;\;
\hbox{invariant}\;\; \hbox{terms}.\eqno{(6.5)}$$ Notice that the
Einstein-Infeld-Hoffmann (EIH) equations \cite{M} can also be
written in the same form (6.5), and hence the conclusion that {\it
the equations (6.4) differ from the EIH
equations by Lorentz invariant terms of order} $c^{-2}$. The
different nature of the coordinate systems for the equations (6.4)
and EIH equations does not permit them to be identical.

Now let us consider a special case of two bodies. From (6.4) we can
determine the relative orbit and then we can compare it with the
corresponding GR relative orbit via EIH equations. It is interesting
that {\em assuming the change of time $d\bar{t}=\Bigl (1+\frac{3}{2}
\frac{G(M+m)}{Rc^2}\Bigr )dt$, the relative orbit from the equations
(6.4) maps into the relative orbit according to the GR}. Since the
reparameterization of the trajectories by the time does not change
the "spatial trajectories", now we generalize the results about the
periastron shift, because now the coordinate system may not rest at
the barycenter of the two bodies and they may not move in a single
plane.

\vspace{-0.2cm}
\section{Some Remarks and Conclusions}\label{sec:9}

Using orthonormal frames enables a deep study on the effects where
angles and precessions are measured and the results regarding the
barycenter of two bodies are the same as in GR. However, not
everything was considered and we complete it now.

\vspace{-0.2cm}
\subsection{The analog of the PPN parameter $\gamma$}

In the GR, the non-zero PPN parameters $\gamma$ and $\beta$ are both
equal to 1. So, it is natural to ask what is their meaning in this
approach. The formulae for the deflection of the light rays,
geodetic precession and the frame dragging effect lead to the
following conclusion: The same formulae which are obtained for
$\gamma =1$ in GR, here are obtained via the equations of motion
(2.23). But, if we omit the matrix $P$ (and $P^T$ also) in the
equations (2.23), then we obtain formulae which are identical for
$\gamma =0$ in the PPN approach, for example in deflection of the
light rays near the Sun, geodetic precession and the frame dragging.
Hence, {\it the appearance of the
matrix $P$ in (2.23) corresponds to} $\gamma =1$. The previous
statement about $\gamma =0$ and $\gamma =1$ can be verified directly
from the equations of motion. The Lorentz force shows that for the
acceleration of a charged particle one should know only the tensor
of electromagnetic field, which is analogous to $\phi_{ij}$, and it
is not necessary to know the velocity of the source and the tensor
$P$ has no role there. Thus, we can intuitively say that $\gamma =0$
in the electrodynamics. Now we clearly see the similarity and the
differences between the electrodynamics and gravitation. Notice that
the Larmour's theorem suggests connection between magnetic field and
the angular velocity, hence simultaneously with the electromagnetic
tensor it is natural to introduce and to consider the tensor $\phi$.

The Coriolis force $\vec f=2m\vec{v}\times \vec{w}$ is obtained in
the equations (3.2) where the tensor $P$ has the essential role. By
omitting the matrix $P$ in (2.23), the result would be $\vec
f=m\vec{v}\times \vec{w}$. Thus, we can write $\vec f=(1+\gamma
)m\vec{v}\times \vec{w}$. The previous discussion and also the
Larmour's theorem are the reason why in many formulae comparing the
angular velocity with the magnetic field, the coefficient 2 appears.

\vspace{-0.2cm}
\subsection{Gravitational radiation}

Further, let us discuss the gravitational radiation. The intensity
of the quadrupole electromagnetic radiation is given by \cite{L}
$$I=\frac{1}{180c^5}\Bigl (\frac{d^3D_{\alpha \beta}}{dt^3}
\Bigr )^2.\eqno{(7.1)}$$ Having in mind that the intensity of the
electromagnetic radiation is proportional to $H^2$ and, $H\sim
\frac{1}{2}w$, we see that in case of gravitation the corresponding
intensity should be $2^2=4$ times larger (assuming a system of units
where $G=1$), i.e., it should be
$$I=\frac{G}{45c^5}\Bigl (\frac{d^3D_{\alpha \beta}}{dt^3}\Bigr )^2.
\eqno{(7.2)}$$ This formula (7.2) is well known for the
gravitational radiation \cite{L} according to the GR. Moreover, it
is known \cite{L} that if the charges of the particles in one system
are proportional to the corresponding masses of the particles, then
there will not exist a dipole electromagnetic radiation. In case of
gravitation, this means that there will not exist dipole
gravitational radiation in case of two bodies.

\vspace{-0.2cm}
\subsection{The analog of the PPN parameter $\beta$}

Analogously to the PPN parameter $\beta$, in flat Minkowski space
we determine a parameter $\beta ^*$ via the expansion of the
coefficient $\mu$:
$$\mu =1+\frac{GM}{rc^2}+\beta ^*\Bigl (\frac{GM}{rc^2}\Bigr )^2+\ldots .
\eqno{(7.3)}$$ Now one can calculate that the perihelion shift is
given by
$$\Delta \varphi = (6\gamma +2\beta ^*){\frac{GM\pi}{ac^2(1-\epsilon^2)}
},\eqno{(7.4)}$$ and since $\beta^*=0$, the total perihelion shift
is a consequence of appearance of the tensor $P$. Comparing this
formula with the corresponding PPN formula, we see that the PPN
parameter $\beta $ corresponds to $2-\gamma +\beta ^*$, which shows
that the tensor $P$ has influence not only on $g_{11}$, $g_{22}$,
and $g_{33}$, but also to $g_{44}$.

\vspace{-0.2cm}
\subsection{Lagrangian}

In section 3 the energy of a particle which moves in a gravitational field
was derived, and it was given by the Lorentz invariant form (3.4d).
Assume that we have a source of gravitation at the coordinate origin, which
rests with respect to the chosen coordinate system. Then the
Lagrangian is given by
$$L=-mc^2\sqrt{1-\frac{v^2}{c^2}}+mc^2\ln \Bigl (1+\frac{GM}{rc^2}\Bigr ).$$
Indeed, a direct calculation shows that the Hamiltonian function
is given by
$${\cal H}=\vec{v}\frac{\partial L}{\partial \vec{v}}-L=
\frac{mc^2}{\sqrt{1-\frac{v^2}{c^2}}} -mc^2 \ln \Bigl (1+\frac{GM}{rc^2}\Bigr )$$
and it is a constant according to (3.4c). Further the Euler-Lagrange
equations can be written in the following equations
$$\frac{dV_i}{ds}=\phi_{ij}V_j,\eqno{(7.5)}$$
for $i=1,2,3$. Compared with the equations (2.23) we notice that if we dismiss
the tensor $P$ and also $P^T$, we obtain (7.5). On the other side, we mentioned
in section 3, that these equations (7.5) lead to the same Hamiltonian function.
Indeed, the appearance of the tensor $P$ means presence of a force,
which does not do action, i.e. preserves the
energy of the test body in the gravitational field. Finally, notice that
analogously to (3.4d),
the Lagrangian can be written in the following Lorentz invariant form
$$L=-\frac{mc^2}{U_iV_i}+mc^2\ln \Bigl( 1+\frac{GM}{rc^2}\Bigr ),\eqno{(7.6)}$$
where $U_i$ is the 4-vector of velocity of the gravitational body, $V_i$ is the
4-vector of velocity of the test body with negligible mass $m$ and
the distance $r$ is determined in the system where the gravitational body rests.

\vspace{-0.2cm}
\subsection{The field equations}

According to the discussion in 2.2, the gravitational potential in case of
many distinct bodes with point masses is given by
$$\mu =\prod_{i}\Bigl (1+\frac{Gm_i}{r_ic^2}\Bigr)\eqno{(7.7)}$$
where the distance $r_i$ between the considered test body and the
$i$-th gravitational body is determined in the system where the $i$-th
gravitational body rests. The scalar $\mu$ can be interpreted as a scalar
which is related to the gravitational redshift caused by many bodies.

In case of a mass distribution given by the mater density $\rho$,
the potential $\mu$ is given by
$$\ln \mu =\int_{V'} \ln
\Bigl (1+\frac{G\rho (\vec{r}')}{\vert \vec{r}-\vec{r}'\vert c^2}dV'\Bigr),
\eqno{(7.8)}$$
where the distance $\vert \vec{r}-\vec{r}'\vert$ is defined analogous as in
(7.7).

Now the density $\rho$ satisfies the following
partial differential equations
$$\frac{\partial\phi_{ij}}{\partial x_k}+\frac{\partial\phi_{jk}}{\partial x_i}+
\frac{\partial\phi_{ki}}{\partial x_j}=0\eqno{(7.9)}$$
and
$$\frac{\partial\phi_{ij}}{\partial x_j}=\frac{4\pi G\rho}{c^2}U_i\eqno{(7.10)}$$
where $U_i$ is the field of 4-vector of velocity of the mater distribution.
They are the field equations and
they are completely analogous to the Maxwell's equations for electrodynamics.
If we put $i=4$ and $u\approx 0$ in (7.10), we just obtain the Poisson's equation.

\vspace{-0.2cm}
\subsection{Short discussion about the Einstein-Infeld-Hoffmann equations}

At the end we try to explain why EIH equations correspond to the
case 4 in section 1. There are two main reasons: i) While there is
no privileged metric in any metric gravitational theory for
determining any (invariant) scalar, for example curvature scalar,
for calculating some noninvariant scalars we really need a
privileged system. For example, we apply the equations of motion
from any metric theory to find many scalars (like perihelion shift
per orbit) assuming a priori that the corresponding equations are
related to a flat manifold, but not curved. These calculations may
lead to satisfactory results, which really happens, only in an
inertial system, i.e. far from the massive bodies.  
ii) The
fact that the equations (6.4) differ from the EIH equations for
Lorentz invariant terms also suggests that both equations are given
with respect to an observer far from massive bodies. 
The equations of
motion according to an observer inside the gravitational field are
much more complicated \cite{TC3}.

\end{document}